\documentclass[final,superscriptaddress,english,twocolumn,amssymb, nobibnotes, aps, prb, longbibliography]{revtex4-1}
\usepackage{graphicx}
\usepackage{natbib}
\usepackage{amsmath}
\usepackage{amssymb}
\usepackage{appendix}
\usepackage[dvipsnames]{xcolor}
\usepackage[section]{placeins}
\definecolor{darkblue}{rgb}{0,0,.65}
\definecolor{darkgreen}{rgb}{0.28,0.41,0.19}

\usepackage[%
    pdfauthor={David J. Luitz},%
  pdfstartview=FitH,%
  breaklinks=true,%
  bookmarks=true,%
  colorlinks=true,%
  anchorcolor=black,%
  citecolor=blue,
  filecolor=black,%
  menucolor=black,%
  urlcolor=darkblue,%
  linkcolor=blue,%
 ]{hyperref}
\usepackage[all]{hypcap} %
\newcommand{\bra}[1]{\langle\,#1\,|}
\newcommand{\ket}[1]{|#1\rangle}

\newcommand{\braket}[2]{\langle\,#1\, | \, #2\,\rangle}

\newcommand{\D}{\mathrm{d}}

\newcommand{\nh}{non-hermitean }

\definecolor{nicegreen}{RGB}{0,170,0}

\begin{document}

\title{Exceptional points and the topology of quantum many-body spectra}

\author{David J. Luitz}
\email{dluitz@pks.mpg.de}
\author{Francesco Piazza}
\email{piazza@pks.mpg.de}
\affiliation{Max Planck Institute for the Physics of Complex Systems, Noethnitzer Str. 38, Dresden, Germany}
\date{June 5, 2019}

\begin{abstract}

We show that in a generic, ergodic quantum many-body system the interactions
induce a non-trivial topology for an arbitrarily small non-hermitean
component of the Hamiltonian. This is due to an
exponential-in-system-size proliferation of exceptional points which
have the hermitian limit as an accumulation
(hyper-)surface. 
The nearest-neighbour level repulsion characterizing
hermitian ergodic many-body sytems is thus shown to be a projection of
a richer phenomenology where actually all the exponentially many pairs of
eigenvalues interact. The proliferation and accumulation of
exceptional points also
implies an exponential difficulty in isolating a local ergodic quantum
many-body system from a bath, as a robust topological signature remains
in the form of exceptional points arbitrarily close to the hermitian limit.
\end{abstract}
\maketitle

\section{Introduction}

The discovery of topological phases has provided a major
paradigm-shift in the understanding of quantum
states \cite{Ryu_2016}. In the case where the system is described by a hermitean Hamiltonian, the interplay of topology with interactions in
many-body systems is now a major avenue of research \cite{Hohenadler_2013,Rachel_2018}.
Currently, a lot of attention is being devoted to
the classification of topological phases also for non-hermitean
Hamiltonians
\cite{PhysRevA.87.012118,PhysRevLett.118.040401,PhysRevB.95.184306,shen_topological_2018,Kunst_2018,Gong_2018,Martinez_Alvarez_2018,PhysRevLett.121.086803,Xiong_2018,Ghatak_2019}
in one
\cite{PhysRevLett.102.065703,Yuce_2015,PhysRevLett.115.200402,PhysRevLett.116.133903,San_Jose_2016,PhysRevB.95.174506,PhysRevA.97.052115,PhysRevB.97.045106,PhysRevB.97.121401,PhysRevA.98.042120,PhysRevB.98.115135,PhysRevB.99.201103}
and higher \cite{PhysRevB.84.205128,Xu_2017,Carlstr_m_2018,PhysRevLett.120.146601,Yao_2018,PhysRevB.98.035141,PhysRevB.99.041406,PhysRevB.99.081102,PhysRevB.99.075130,mcclarty2019non} spatial dimensions, with the most direct physical application being to dissipative
systems \cite{Rotter_2009}, where controlled experimental platforms
are already available \cite{PhysRevLett.104.153601,Brandstetter_2014,Gao_2015,Zeuner_2015,Poli_2015,Doppler_2016,Xu_2016,Peng_2016,Chen_2017,Bandres_2018,Zhou_2018,1808.09541v2}. This task is still in progress for single-particle
bands, so that the role of interactions in quantum many-body \nh 
systems remains largely unexplored in relation to topology.

In all the above explorations, the non-trivial topology, the interactions,
and the \nh nature are ingredients which can be separately
added to the picture. Here we introduce a generic scenario where the above three elements are
instead deeply interconnected. We show that in an ergodic quantum
many-body system the interactions induce a non-trivial topology for an \emph{arbitrarily small} \nh component of the
Hamiltonian. This is due to an exponential-in-system-size
proliferation of exceptional points which have the hermitian
limit as an accumulation (hyper-)surface. Exceptional points are known to carry
non-trivial topological features \cite{heiss_physics_2012} and indeed play a crucial role in the
understanding of topological phases in \nh bands. The connection
between level repulsion in the Hamiltionan spectrum of an ergodic
system and the distribution of exceptional point has been so far
argued based on toy models \cite{heiss_avoided_1990} and
demonstrated at fine-tuned points of models without local degrees of freedom which
become classical in the thermodynamic limit
\cite{Heiss_2005,Cejnar_2018}.
Here we demonstrate such scenario in \emph{fully
generic local many-body systems with no semiclassical limit and
without any fine-tuning of microscopic parameters}.

A further remarkable and
defining feature of generic ergodic many-body system follows from
the exponential proliferation of exceptional point arbitrarily close to the
hermitean limit: The many-body spectrum can be understood as a single
Riemann surface, where all eigenvalues are adiabatically connected along smooth
  paths, as any of the $N(N-1)/2$ possible pairs of
eigenvalues can be interchanged by encircling the corresponding exceptional points. Here $N=\text{dim}(\mathcal H)$ is the dimension of
the Hilbert space which grows exponentially with the system size $L$ of the
quantum many-body system.

Our results suggest a deeper connection than so far expected between topology and
interactions, and also establish a new way of thinking about generic
quantum many-body systems.

A general physical implication of our findings is the resulting \emph{exponential-in-sistem-size difficulty of isolating an ergodic quantum many-body system from an external
bath}, as a topologically robust signature of the openness always remains
in the form of exceptional points.

\section{Model}
\label{sec:model}

\begin{figure}[h]
	\includegraphics[width=\columnwidth]{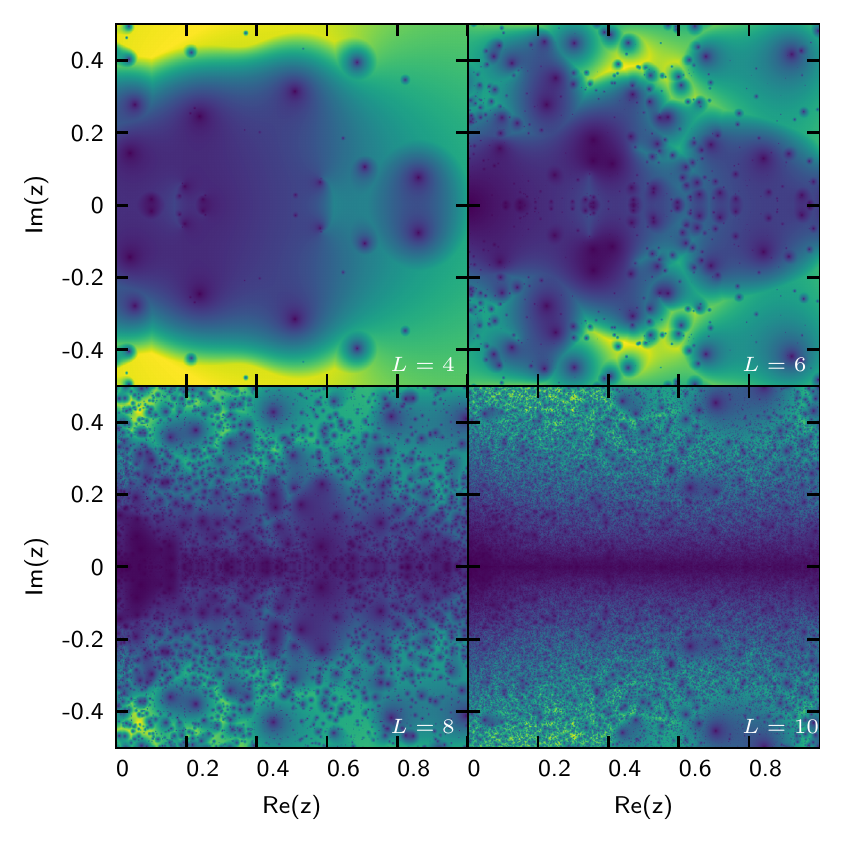}
	\caption{Minimal eigenvalue distance $\min\limits_{i,j} \left| \lambda_i(z) - \lambda_j(z) \right|$ between all eigenvalues $\lambda_i(z)$ of the Hamiltonian \eqref{eq:Ham} as a function of the complex parameter $z$ for different lengths $L$ of the spin chain. Minima (dark colors) correspond to degeneracies of the spectrum, which we identify to be exceptional points.
		\label{fig:mindist}
	}
\end{figure}

In order to demonstrate the above scenario we consider
a quantum spin $\frac 1 2$ chain, described by the Hamiltonian
\begin{equation}
\begin{split}
\hat H(z) &= \sum_{i=1}^{L-1} J_x \hat \sigma_i^x \hat \sigma_{i+1}^x + J_y \hat \sigma_i^y \hat \sigma_{i+1}^y + J_z \hat \sigma_i^z \sigma_{i+1}^z  \\
&+ \vec{h_1}\cdot \hat \vec{\sigma_1} + \vec{h_L} \cdot \hat \vec{\sigma_L}\\
&+ z \sum_{i=1}^L \left( g_{xz} \hat \sigma_i^x \hat \sigma_{i+1}^z + g_{xy} \hat \sigma_i^x \hat \sigma_{i+1}^y + g_{yz} \hat \sigma_i^y \hat \sigma_{i+1}^z \right),
\end{split}
\label{eq:Ham}
\end{equation}
with fixed real parameters $J_x, J_y, J_z, g_{xz}, g_{xy}, g_{yz} \in \mathbb{R}$ and fixed real boundary fields $\vec{h}_1, \vec{h}_L \in \mathbb{R}^3$. The site indices are defined modulo $L$, where needed (last sum in Eq. \eqref{eq:Ham}). With these choices, the Hamiltonian is a matrix valued function of one scalar complex parameter $z\in \mathbb{C}$. 
For all $z$ on the real axis, the Hamiltonian is hermitean, while it is non-hermitean in general.
Throughout this paper we take the generic choice of parameters $J_x=1.2$, $J_y=1.0$, $J_z=0.7$, $g_{xz}=0.91$, $g_{xy}=0.7$, $g_{yz}=1$, $\vec{h}_1 = (0.0291241 , 0.02341097 , 0.0567)^T$, $\vec{h}_L = (0.091241 , 0.018924 , 0.0781652)^T$, such that the Hamiltonian has no symmetries. 
Note that this parameter set is not fine tuned and that we find the
same phenomenology with other parameter choices. 

We have verified that the Hamiltonian for $z\in\mathbb{R}$ is ergodic in the sense that local observables thermalize by means of the eigenstate thermalization hypothesis\cite{deutsch_quantum_1991,srednicki_chaos_1994,rigol_thermalization_2008,dalessio_quantum_2016,borgonovi_quantum_2016}, which is valid in this system (cf. supplementary material in Sec. \ref{sec:ergodicity}). Furthermore, we have considered the statistics of level spacings of the hermitean Hamiltonian, using the ratio of adjacent gaps\cite{oganesyan_localization_2007}. We find that the spectral statistics are in the gaussian unitary ensemble (GUE) universality class (cf. Sec. \ref{sec:ergodicity}).

\section{Proliferation of exceptional points}

\begin{figure}[h]
	\includegraphics[width=\columnwidth]{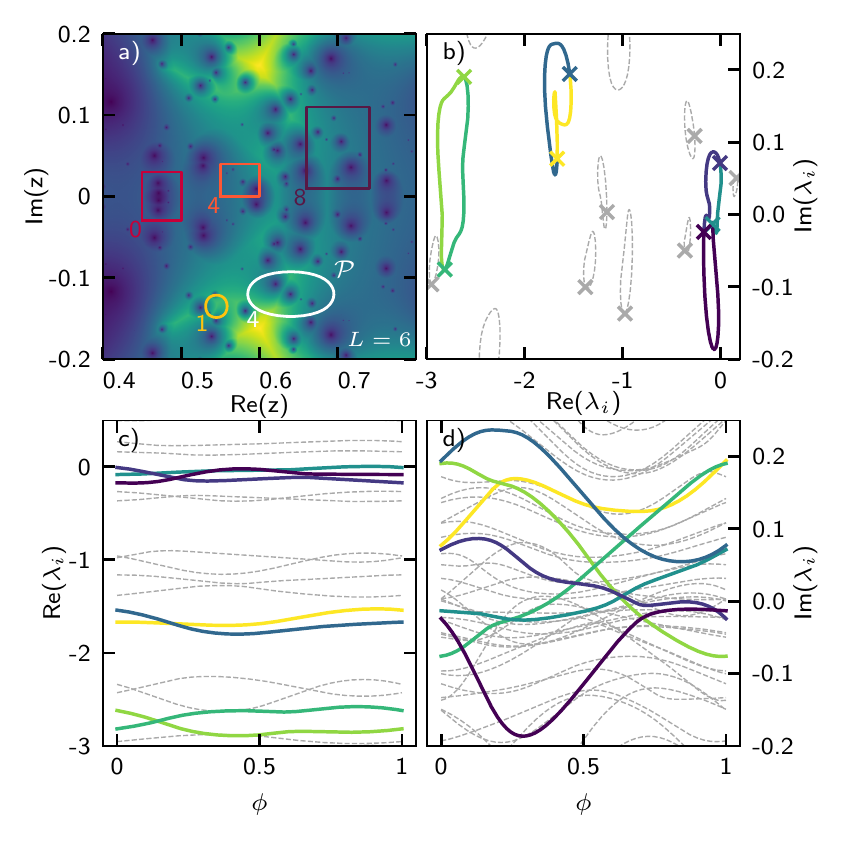}
	\caption{Braiding of eigenvalues $E_n(z)$ along closed paths
		$g(\phi)$ encircling exceptional points for a chain of
		length $L=6$. a) Overview of minimal eigenvalue distances
		$\min_{i,j}(|\lambda_i-\lambda_j|)$ as a function of the
		complex parameter $z$. Closed curves are examples with the
		number indicating the number of transpositions in the
		permutation linking the final eigenvalues $\lambda_i(1)$
		with their initial order $\lambda_i(0)$. b) For the white path $\mathcal{P}$ in panel a), we show the evolution along the curve of the relevant eigenvalues (colored). Lower panels: Eigenvalue evolution (real (c) and imaginary (d) part) $\lambda_i(\phi)$ along $\mathcal{P}$.
		\label{fig:braiding}}
\end{figure} 

Since exceptional points necessarily are related to a degeneracy of
(at least) two eigenvalues of $H(z)$, it is natural to consider the
distances between all eigenvalues $\lambda_i$ of $H(z)$. We show in
Fig. \ref{fig:mindist} the minimal distance of all pairs of
eigenvalues $\delta(z) = \min_{i,j} \left| \lambda_i(z) - \lambda_j(z)
\right|$ as a function of the complex interaction parameter $z$. If this distance $\delta(z)$ vanishes, it corresponds to a degeneracy of at least two eigenvalues. While $\delta(z)$ is a continuous function of $z$, it is not analytic, and exhibits a large number of kinks, when the closest pair of eigenvalues changes.

We observe that as a function of system size the number of very small
eigenvalue distances increases significantly. It is also clear that 
the minima of $\delta(z)$ appear to be very sharp, consistent with the
typical square root singularity of exceptional points. The extreme proliferation of degeneracies seems to occur most strongly close to the real axis, when $\text{Im}(z)=0$.

While the results of Fig.~\ref{fig:mindist} are already a strong
indication of the proliferation of exceptional points and their accumulation on the
real axis $\text{Im}(z)=0$, in order to have solid quantitative
characterization we need to resort to the defining feature of exceptional points: the occurrence of a square root branching point,
distinguishing them from other possible degeneracies. This leads to the swapping of a pair (or more in the case of higher order exceptional points, which we do not observe here) of eigenvalues when following a closed path round an exceptional point. We will extensively use this property to study the density of exceptional points with respect to the distance from the real axis in parameter space.

\section{Braiding on a single Riemann sheet}

In Fig. \ref{fig:braiding} we demonstrate that the degeneracies
appearing in Fig.~\ref{fig:mindist} are indeed associated with exceptional points. Panel a) contains a colormap of the minimal eigenvalue distance $\min_{i,j}(|\lambda_i - \lambda_j|)$ as in Fig. \ref{fig:mindist} for a chain of length $L=6$. The location of degeneracies in the spectrum is clearly visible in dark spots. %
The rectangluar and elliptical paths drawn in Fig. \ref{fig:braiding}
a) correspond to example paths we consider, encircling different
numbers of exceptional points. Next to each path, we list the number of
transpositions in the permutation of eigenvalues after closure of the
path. This is achieved by tracking the eigenvalues along the path as
described in Appendix \ref{sec:tracking} and afterwards analyzing the resulting
permutation as described in Appendix \ref{sec:counting}. Within one half plane of the complex plane, these numbers correspond to the number of encircled exceptional points, while exceptional points at conjugate locations ($z$ and $z^*$) have opposite handedness and therefore undo swaps mutually. 
We note here that in our system each exceptional point has a conjugate partner, since $H(z^*)=H(z)^\dagger$, and the spectra of $H$ and $H^\dagger$ are identical (since this corresponds to the adjoint eigenvalue problem).
The swapping of eigenvalues (crosses) $\lambda_i$ is exemplified in
Fig. \ref{fig:braiding} b), where each colored line corresponds to the
trajectory in the complex eigenvalue space of one eigenvalue along one
traversal of the path labelled by $\mathcal{P}$ in Fig. \ref{fig:braiding} a). It is apparent that the four exceptional points enclosed by the path swap two pairs of eigenvalues, and permute three more eigenvalues in a cycle corresponding to two transpositions.

Any conjugate pair of exceptional points in the complex parameter plane is connected
by a branch cut, interconnecting the Riemann sheets on which each
eigenvalue evolves. Due to the
proliferation of exceptional points, the Riemann surface of our
ergodic quantum many-body model becomes massively interconnected. As
we shall see next, the proliferation is exponential in the system
size $L$ such that there is one exceptional point for each possible pair of eigenvalues.
This means that starting from one eigenvalue $\lambda_i$, any other
eigenvalue can be reached by adiabatic parameter changes in the
complex plane, that is, the spectrum of an ergodic quantum many-body
system actually belongs to a \emph{single} Riemann surface.

\section{Statistics of exceptional points}

\begin{figure}[h]
	\includegraphics[width=\columnwidth]{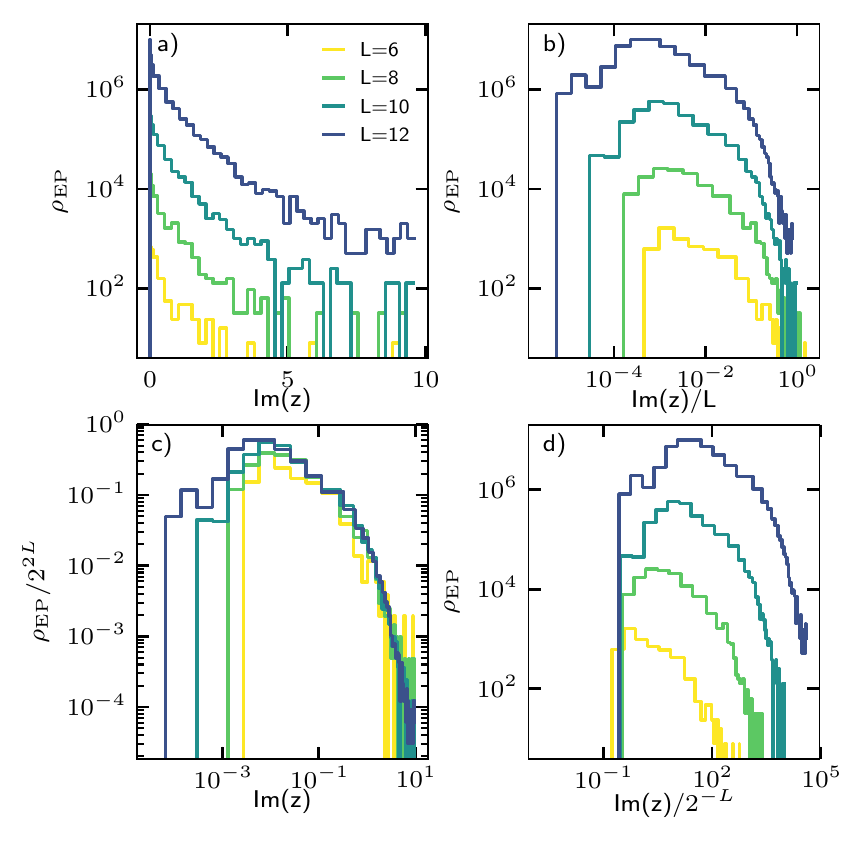}
	\caption{Density $\rho_{\rm EP}$ of exceptional points per unit area in the complex parameter plane. We consider parameter areas at $\text{Re}(z)=0.5$ and count the number of exceptional points in areas of size $\D \text{Re}(z) \D \text{Im}(z)$, with exponentially decreasing bin size $\D \text{Re}(z)\propto 2^{-L}$. a) Density versus distance from the real axis in parameter space for different system sizes. b) Density versus distance from the real axis rescaled by $1/L$. c) Density rescaled by the total number of exceptional points $2^{2L}$. d) Density versus distance rescaled by $1/2^{-L}$.
		\label{fig:ep_distribution}}
\end{figure}

\begin{figure}[h]
	\includegraphics[width=\columnwidth]{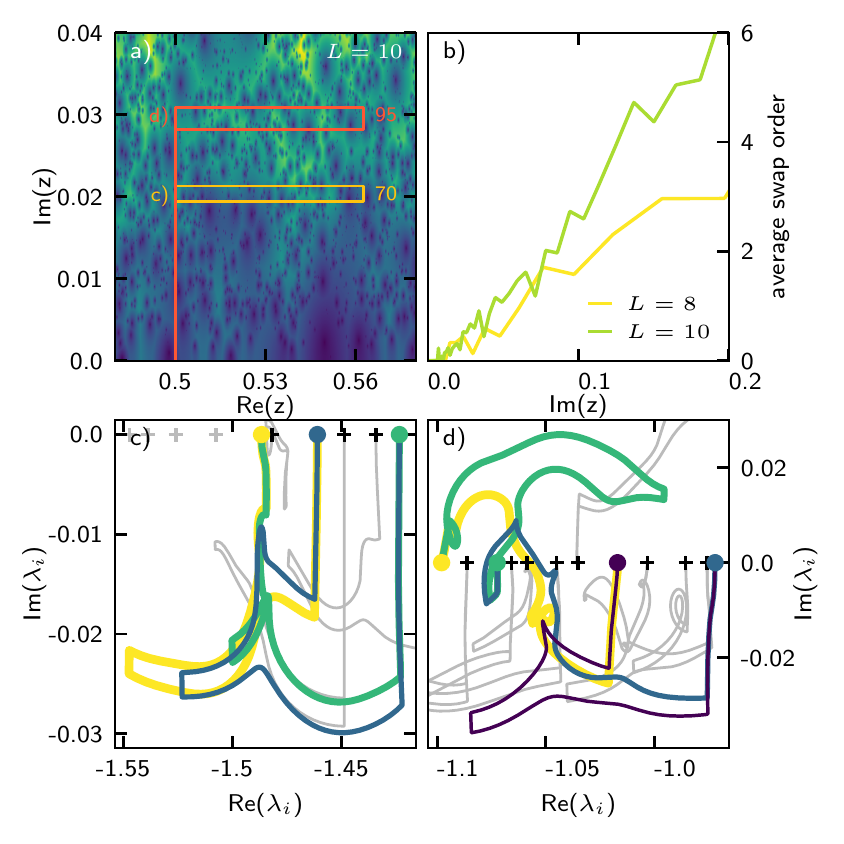}
	\caption{a) Zoomed overview of the minimal eigenvalue distance $\min_{i,j} |\lambda_i - \lambda_j|$ for a system of size $L=10$. Solid lines are the two exemplary paths connected to the real parameter axis for which eigenvalue traces are shown in c) and d).
		b) Average distance of swapped eigenvalues of the hermitean system by exceptional points as a function of their distance from the real axis for $L=8$ and $L=10$. c) and d) exemplary eigenvalue cluster traces swapped (colored) by exceptional points encircled by the paths shown in panel a).
		\label{fig:higher-order-braiding}}
\end{figure}

To characterize the proliferation of exceptional points with increasing
system size, we compute their distribution in the complex plane $\mathrm{Re}(z), \mathrm{Im}(z)$. In
Fig.~\ref{fig:ep_distribution} we show the density of exceptional points as a
function of the distance from the real axis $\mathrm{Im}(z)$. We
compute the density $\rho_{\rm EP}=M/A$
by counting the number $M$ of exceptional points within a given area $A$ in the
complex plane (as described above). In order to collect enough statistics still keeping
computational times feasible, we choose the area $A$ such that it
typically contains between 100 and 1000 exceptional points in the bulk of the distribution, independently of the system's size
(which as we shall see implies exponential down-scaling of $A$ with $L$).
As apparent from Fig.~\ref{fig:mindist}, the density becomes more and more
independent of the position on $\mathrm{Re}(z)$ as $L$ is
increased. Selecting areas $A$ of a finite width along $\mathrm{Re}(z)$, we effectively average over $\mathrm{Re}(z)$ to improve the statistics.

Panel a) shows the density up to large distances from the real axis,
so that the tails of the distribution (compatible with
an exponential decay) are visible. More interesting is the distribution in the
vicinity of the real axis, as shown in panels b) to d). In panel b),
we see that the distribution extends up to distances from the real
axis which scale with the system size $L$. Moreover, as apparent from panel c), the
overall scale of the density increases exponentially as $2^{2L}$. Taking into account the fact
that the tails of the distribution extend to values of
$\mathrm{Im}(z)$ of the order of $L$, this scaling 
is consistent with having a proliferation of exceptional points such
that we have one for each possible pair of eigenvalues of the
Hamiltonian \eqref{eq:Ham} i.e. $N(N-1)/2$ with $N=\text{dim}(\mathcal
H)=2^L$. Finally, panel d) shows that the bottom of the distribution
has a gap from the real axis which vanishes like $2^{-L}$, which
demonstrates the exponentially fast accumulation of exceptional points at the
hermitean line $\mathrm{Im}(z)=0$. This scaling of the gap is to be
expected under two assuptions: i) the exceptional points at the bottom of the distribution are the
ones connected to the avoided crossings between nearest-neighbor levels of the hermitean Hamitonian
$\mathrm{Im}(z)=0$ (see discussion below and
Fig.~\ref{fig:higher-order-braiding}); ii)
the distance of each exceptional point from the real axis is proportional to the real
gap characterizing the corresponding avoided level
crossing. Assumption ii) is actually a mathematical fact for a single
exceptional point but not trivial in presence of multiple exceptional points, which are known to
influence each-other strongly. This issue, which has been investigated
in toy models \cite{heiss_avoided_1990} and fine-tuned semiclassical
models without local degrees of freedom \cite{Heiss_2005,Cejnar_2018}, is even less
trivial for our quantum many-body model showing an exponential
proliferation of exceptional points.
Since the number of avoided level crossings in the
hermitean case $\mathrm{Im}(z)=0$ scales as $2^L$ and the spectral
bandwidth only as $L$, the corresponding real gaps must scale down
exponentially like $2^{-L}$. Therefore, making the assumptions i) and
ii) we can conclude that the exceptional points at the bottom of the distribution
approach the real axis like $2^{-L}$.

The assumption i) can actually be tested by analyzing the braidings
between eigenvalues along closed paths taken at different distances
from the real axis. The result of this analysis is shown in
Fig.~\ref{fig:higher-order-braiding}. Taking paths which start and end
at the real axis we can attribute a given order to the swap generated by
an exceptional point. As illustrated by panels c) and d), this is given
by the number of real eigenvalues of the initial hermitean Hamiltonian, which lie between the two which get
swapped. For large system sizes such a path encircles so many exceptional points that
only the analysis of the permutation of the levels after closure
of the path remains possible. As described in section \ref{sec:swap_order}, from the cycle-structure of the permutation we can
extract an average swap order as a function of the distance from the
real axis, shown in panel b). The bottom of the distribution of exceptional points shown in
Fig.~\ref{fig:ep_distribution} lies well within the region where the
average swap order is well below 2. Panel b) also indicates a linear
growth of the average swap order as a function of the distance from
the real axis. This supports the hypothesis that assumption ii) above not
only applies to nearest neighbour avoided level crossings but also to
eigenvalues which are not neighbours in the hermitean limit.
Indeed, applying this assumption to a pair of levels in the hermitean spectrum
which are separated by a given number of levels $k=0,\dots,2^L-2$, we
would conclude that the typical distance from the real axis of an exceptional point swapping such
levels would scale as $d_k\sim (L/2^L) k$, since the spectral bandwidth of
the hermitian Hamiltonian scales as $L$ and the typical level
separation scales like $2^L$. This argument reproduces the linear
scaling observed in Fig.~\ref{fig:higher-order-braiding}b). The
estimated slope
however does not agree with our numerics, probably because this rough argument  does not
take into account the inhomogeneity of the density of states.
The above scaling expression for $d_k$ also predicts the exceptional point
distribution to extend up to values of
$\mathrm{Im}(z)$ scaling with $L$, in agreement with Fig.~\ref{fig:ep_distribution} b).

The picture that emerges from these findings is that, while increasing
the system size $L$, exceptional points generating a
braiding with swapping order $k\ll 2^L$ approach the real axis
exponentially fast. The fastest class corresponding to the
nearest-neighbour swaps $k=0$, defining the bottom of the exceptional point
distribution.

\section{Discussion and Outlook}

The results presented here provide a way of characterizing ergodicty in
quantum many-body systems which was so far unexplored in the field of condensed
matter physics.

One interesting physical implication concerns the interaction between
the levels. In the purely hermitean picture these interactions take
place pairwise between nearest-neighbours, which underlies the
emergence of random Gaussian spectral statistics and the eigenstate
thermalization hypothesis (see discussion at the end of section
\ref{sec:model}). Here we verify -- for the first time to the best of our knowledge in a generic quantum many-body
system -- the hypothesis that this scenario is a
projected manifestation of the more complex phenomenology of
eigenvalue-braiding through exceptional points in the complex plane. More
interestingly, we show that \emph{exponentially close to the hermitean
limit} actually not only nearest-neighbour levels interact via exceptional points, but also
all levels which are separated by a number of levels which is not
exponentially large in system size.

A further (and even more physically transparent) implication emerging from
our findings is that the difficulty of isolating an ergodic
quantum many-body system from its environment is exponentially large in system's
size. By this we mean that the coupling strength between the ergodic system
and the environment needs to be exponentially small in order for the
system not to be affected by the exceptional points. Indeed, the \nh component of
the Hamiltonian is generated by tracing out the environment and is
thus proportional to the strength of the coupling to the latter.
It is important to note at this point that our results are indeed directly applicable to open
systems in contact with an environment despite the fact that they consider only a \nh Hamiltonian
without the corresponding bath-induced fluctuations
\cite{gardiner2004quantum}. The reason is that we are interested in
the exceptional points, and these proliferate even for exponentially small \nh component, where
the damping as well as the corresponding noise are exponentially
small, so that only the topological signature of the exceptional point is left.

Apart from suggesting a deep connection between topology and
interactions which seems worth pursuing further, the present findings and
the above implications open a new perspective on many-body quantum ergodicity which is also very timely, as there is currently a large
interest in studying the effect of the openness on
generic scenarios for relaxation and thermalization\cite{levi_robustness_2016,luschen_signatures_2017}.

Several lines of further investigation emerge naturally out of the
present work, like the analysis of spectral statistics in the
complex plane and its connection to exceptional points, the investigation of the
consequences of the exceptional points-proliferation for the dynamics, as well as the
extension of the present study to disordered and driven systems.

\acknowledgements

We would like to thank Roderich Moessner, Frank J\"ulicher, Sebastian
Diehl, Paul McClarty, and Jeff Rau for helpful discussions.
We furthermore acknowledge PRACE for awarding access to 
HLRS’s Hazel Hen computer based in Stuttgart, Germany under grant number 2016153659.

\clearpage

\appendix

\section{Ergodicity of the hermitean model}
\label{sec:ergodicity}

\begin{figure}
	\includegraphics[width=\columnwidth]{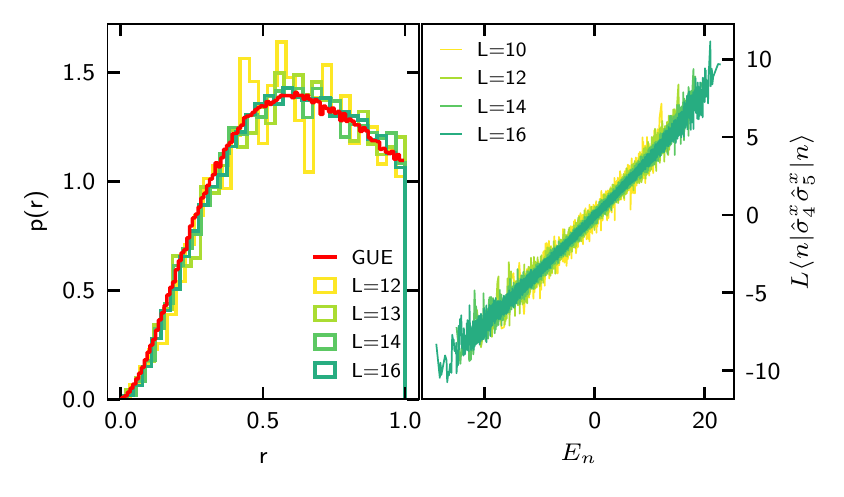}
	\caption{Left: Distribution of adjacent energy gap ratios $r_n=\min((E_{n+1}-E_{n})/(E_{n}-E_{n-1}),(E_{n}-E_{n-1})/(E_{n+1}-E_{n}))$ in comparison with the result for random matrices from the gaussian unitary ensemble (GUE). Right: Eigenstate expectation value $\bra{n} \hat \sigma_4^x \hat \sigma_5^x \ket{n}$ as a function of eigenenergy $E_n$ for different system sizes. Both panels show data for the same parameters as used in the main text and real $z=0.5$, for which the Hamiltonian is hermitean.
		\label{fig:ergodicity}
	}
\end{figure}

Here we show additional data to demonstrate that the model with the parameters used in the main text is indeed fully ergodic in the hermitean limit.
We consider in Fig. \ref{fig:ergodicity} two aspects of ergodicity: In the left panel, we compare the statistics of the ratio of adjacent energy levels\cite{oganesyan_localization_2007} $r_n=\min((E_{n+1}-E_{n})/(E_{n}-E_{n-1}),(E_{n}-E_{n-1})/(E_{n+1}-E_{n}))$ for different system sizes $L$ with the result from random matrix theory in the gaussian unitary ensemble (GUE), the distributions match very well and strong level repulsion is visible as predicted in random matrix theory.

The second criterion we use to quantify ergodicity is the validity of the eigenstate thermalization hypothesis\cite{deutsch_quantum_1991,srednicki_chaos_1994,rigol_thermalization_2008,dalessio_quantum_2016,borgonovi_quantum_2016} (ETH), which ensures thermalization of the closed system. One condition is that eigenstate expectation values of local operators should become a smooth function of the eigenenergy in the thermodynamic limit $L\to \infty$, coinciding with the expectation value of the operator in the microcanonical ensemble. In the right panel of Fig. \ref{fig:ergodicity}, we show results for the operator $ \hat \sigma_4^x \hat \sigma_5^x $, which is a term in the Hamiltonian and a part of the energy density, therefore showing a strong positive correlation with the energy eigenvalue. The results are clearly consistent with the eigenstate thermalization hypothesis. We therefore conclude that the model used in the main text is fully ergodic.

\section{Tracking eigenvalues}
\label{sec:tracking}

To extract the number of swaps of eigenvalues, it is necessary to track the evolution of each eigenvalue along a periodic curve $g(\phi):[0,1]\to \mathbb{C}$ with $g(0)=g(1)$. After closing the curve, the spectra of $\hat H(g(0))$ and $\hat H(g(1))$ are identical, up to a permutation of the eigenvalues, where each eigenvalue is a continuous function of the curve parameter $\phi$.

It is in general a formidable task to track the evolution of eigenvalues of a quantum many-body Hamiltonian $\hat H$ as a function of a (scalar) parameter. Here, we consider a complex parameter $z \in \mathbb{C}$, including the possibility that eigenvalues undergo branch cuts. We are using perturbation theory for this task, which allows us to calculate the derivatives of each eigenvalue with respect to the parameter change along the curve. Comparing the exact new eigenvalues after a small step along the curve with the predictions, we can match each new eigenvalue to the previous ones.
  
Consider the eigenvalues $E^{(0)}_n$ of $\hat H(z_0)$ at a point $z_0$ in the complex plane. We can predict the eigenvalues of $\hat H(z_0+\epsilon)$ for a small (complex) parameter change $\epsilon$ using non-hermitean perturbation theory.

Let $\ket{n_0}$ be the right eigenvectors of $\hat H(z_0)$ with eigenvalue $E_n^{(0)}$.
Let further $\ket{\tilde{n_0}}$ be the left eigenvectors of $\hat H(z_0)$ with eigenvalue $E_n^{(0)*}$. The normalization of the eigenvectors can be chosen such that left and right eigenvectors are orthonormal (note that right/left eigenvectors themselves are in general not orthogonal)\footnote{We note that Lapack routines like \texttt{zgeev} yield a different normalization for the left/right eigenvectors.}:
\begin{equation}
\braket{\tilde{n_0}}{m_0} = \delta_{n,m}.
\end{equation}

Then, perturbation theory for non-hermitean operators \cite{sternheim_non-hermitian_1972} yields:
\begin{equation}
\begin{split}
E_n(z_0+\epsilon) &= E_n(z_0) \\ 
&+ \epsilon \bra{\tilde{n}_0} \hat D(z_0,\epsilon) \ket{n_0} \\
&+ \epsilon^2 \sum_{m\neq n} 
\frac{\bra{\tilde{n}_0} \hat D(z_0,\epsilon) \ket{m_0}\bra{\tilde{m}_0} \hat D(z_0,\epsilon) \ket{n_0} }{E_m-E_n},
\end{split}
\end{equation}
with the perturbation $\hat D(z_0,\epsilon) = \frac{\hat H(z_0+\epsilon) - \hat H(z_0)}{\epsilon}$. Due to level repulsion in our system, the denominators $(E_m-E_n)$ do not lead to problems if the step size $\epsilon$ is small enough. While degeneracies are possible at exceptional points, the sampled points on the curve did not coincide with exact exceptional points in practice for the system sizes we considered.

This approach allows us to predict for each eigenvalue at $z_0$ its
change at $z_0+\epsilon$, taking into account level crossings and
avoided crossings. We fully diagonalize $\hat H(z_0+\epsilon)$ and
compare the eigenvalues to their predicted locations to attach the
labels $n$ to each eigenvalue. This is necessary for two reasons:
firstly, complex eigenvalues do not have a natural ordering and
secondly, any ordering of eigenvalues used for the labeling would
miss swaps and level crossings.

Our procedure depends on a good choice of the step size $\epsilon$ and it is clear that in the proximity of exceptional points very small step sizes have to be used. Therefore, we developed an adaptive method for tracking all eigenvalues along the curve $g(\phi)$, which for each step $g(\phi)\to g(\phi+\delta_\phi)$ checks if the eigenvalues at $g(\phi+\delta_\phi)$ are sufficiently close to the predictions (compared to the distances between the eigenvalues). If this is not the case, instead of one step $\delta_\phi$, two steps of size $\delta_\phi/2$ are carried out. Applying this procedure recursively, the step size is adjusted as needed, allowing a faithful tracking of eigenvalues of very large \nh Hamiltonians, here up to dimensions of $N=4096$.

\section{Counting of exceptional points}
\label{sec:counting}

  In order to quantify the extreme proliferation of exceptional points with system size visible in Fig. \ref{fig:mindist}, we need a method to count the number of exceptional points in a region in the complex parameter space. 
  Since each exceptional point swaps two eigenvalues of $H(z)$ if a closed path round the exceptional point is followed, we can count the number of exceptional points enclosed by a closed path by counting the number of swaps. Concretely, the number of exceptional points enclosed by a curve $g(\phi):[0,1]\to\mathbb{C}$ is obtained by tracking all eigenvalues $E_n(\phi)$ of $H(g(\phi))$ along the curve and comparing the labels of the eigenvalues of the initial $\phi=0$ and the final point $\phi=1$. Since the spectra are identical, they differ only by a permutation $\pi$: $E_n(0) = E_{\pi(n)}(1)$.
  
  The number of exceptional points encircled by the curve is than equal to the number of transpositions in the permutation $\pi$:
  \begin{equation}
	n_\text{EP} = N - n_\text{cycles}(\pi).
  \end{equation}
  
  \section{Obtaining the order of swaps}
  \label{sec:swap_order}

  In Fig. \ref{fig:higher-order-braiding}, we also analyze which eigenvalues are swapped by exceptional points, when following a path which begins and ends on the real axis in parameter space. For $z\in \mathbb{R}$, the eigenvalues $E_n(z)$ of $H(z)$ are real and can be ordered by their magnitude. Following a closed path in the complex parameter plane, we investigate which eigenvalues are swapped by exceptional points as a function of their location relative to the real parameter axis. Since for large systems typical loops $g(\phi)$ encircle always multiple exceptional points, we have to unravel this information from the final permutation of eigenvalues after closing the loop $E_n(0)  = E_{\pi(n)}(1)$. 
  The permutation $\pi$ is decomposed into cycles of indices, such that indices inside the cycle are shuffled around under multiple applications of the permutation (multiple traversals of the loop), while no other indices are involved in this cycle. Therefore, we can analyze each cycle of the permutation separately, as only the involved eigenvalues are exchanged with each other by the exceptional points enclosed by the path. If only nearest neighbor eigenvalues are involved in each circle, we call this swaps of order 0. If on the other hand an exceptional point exchanges eigenvalues separated by $m$ other eigenvalues not in the cycle, we refer to this as a swap of order $m$.

\bibliography{exceptional_points}
\end{document}